\documentclass[12pt]{article}
\usepackage{epsfig}
\newcommand{\be}{\begin{equation}}

\newcommand{\ee}{\end{equation}}
\newcommand{\ben}{\begin{eqnarray*}}
\newcommand{\een}{\end{eqnarray*}}

\newcommand{\un}[1]{\underline{#1}}
\newcommand{\as}{\alpha_s}

\newcommand{\bas}{{\overline\alpha}_s}
\def\eq#1{{Eq.~(\ref{#1})}}
\def\fig#1{{Fig.~\ref{#1}}}

\begin{document}
\title{\vspace*{1cm}{\bf Perturbative Odderon in the Dipole
Model\footnote{\normalsize{We dedicate this work to the memory of Jan
Kwieci\'nski.}}\\[1cm] }} 
\author{
{Yuri V.\ Kovchegov\thanks{e-mail: yuri@phys.washington.edu}~$\, ^1$, Lech Szymanowski\thanks{e-mail: Lech.Szymanowski@fuw.edu.pl}~$\, ^2$ and Samuel Wallon\thanks{e-mail: Samuel.Wallon@th.u-psud.fr}~$\, ^3$ } \\[1cm] 
{\it\small $^1$Department of Physics, University of Washington, Box 351560} \\
{\it\small Seattle, WA 98195, USA }\\[0.5cm]
{\it\small $^2$Soltan Institute for Nuclear Studies, Ho\.{z}a 69} \\ 
{\it\small 00-681 Warsaw, Poland}\\[0.5cm]
{\it\small $^3$Laboratoire de Physique Th\'eorique, Universit\'e Paris XI,} \\ 
{\it\small centre d'Orsay, b\^atiment 211, 91405 Orsay cedex,
France}
\\[1cm]
}

\date{September 2003}
\maketitle

\thispagestyle{empty}

\begin{abstract}
We show that, in the framework of Mueller's dipole model, the
perturbative QCD odderon is described by the dipole model equivalent
of the BFKL equation with a $C$-odd initial condition. The
eigenfunctions and eigenvalues of the odderon solution are the same as
for the dipole BFKL equation and are given by the functions $E^{\,
n,\nu}$ and $\chi (n,\nu)$ correspondingly, where the $C$-odd initial
condition allows only for odd values of $n$. The leading high-energy
odderon intercept is given by $\alpha_{odd} - 1 = \frac{2 \, \as \,
N_c}{\pi} \, \chi (n=1 ,\nu=0) = 0$ in agreement with the solution
found by Bartels, Lipatov and Vacca. We proceed by writing down an
evolution equation for the odderon including the effects of parton
saturation. We argue that saturation makes the odderon solution a
decreasing function of energy. 
\end{abstract}

\thispagestyle{empty}
\begin{flushright}
\vspace{-20cm}
NT@UW--03--025 \\
INT ACK 03--51
\end{flushright}

\newpage

\setcounter{page}{1}

\section{Introduction}

The dominant contributions to the total cross sections of hadronic
reactions are related to the Pomeron and to the Odderon
exchanges. While the Pomeron exchange having the quantum numbers of
the vacuum represents a dominant contribution to the sum of total
cross sections for a given hadronic process and its crossing
counterpart which is even under crossing, the Odderon exchange
dominates the difference of these two cross sections, which is odd
under the crossing symmetry. Within Regge theory Pomeron and Odderon
are therefore natural partners. The importance of Odderon exchange for
the phenomenology of hadronic reactions was noted long ago
\cite{LukNic} and was the subject of many investigations. A
review of these early studies can be found in, e.g., \cite{Nic}.

With the advent of quantum chromodynamics (QCD) theoretical status of
the perturbative Odderon followed the development of the analogous
description of its Pomeron partner.  Within QCD in the leading
logarithmic approximation (LLA), the Odderon appears as the color
singlet exchange of three gluons, which interactions are of a form
very similar to the interactions of the two gluons forming the
Pomeron. It is therefore not surprising that soon after the derivation
of the Balitsky-Fadin-Kuraev-Lipatov (BFKL) equation
\cite{BFKL} for the Pomeron there was derived an analogous 
Bartels-Kwiecinski-Praszalowicz \cite{BJKP} equation for the Odderon.

The solution of the BFKL equation by Lipatov in the seminal Ref.
\cite{Lipatov1} stimulated similar studies of the BKP equation which, 
because of its three-body nature, turned out to be much more difficult
to solve.  These attempts led nevertheless to a discovery of a deep
connection between the BFKL equation, the BKP equation and their
generalizations on one side and the two dimensional conformal models
of spin chains \cite{Lev93} and their exact solubility
\cite{Lev94,Korch95} on another side.  Finally they also led to
discovery of several solutions of the BKP equation \cite{JW},
\cite{KKM}, \cite{BLV} with different intercepts of the corresponding Odderon 
Regge trajectories determining the asymptotics of the cross
sections. In particular, the solution found by Bartels, Lipatov and
Vacca (BLV) in Ref. \cite{BLV} describes the Odderon with the largest
intercept found to date, which is equal to $1$\footnote{Here we
restrict ourselves to the Odderon described by the BKP equation for
three gluons in the $t$-channel. In Ref. \cite{DKKM2} it was argued
that exchanges of many $t$-channel gluons generalizing the BKP
equation may also lead to the Odderon intercept asymptotically
approaching $1$ from below.}. Thus the present results suggest that
the perturbative Odderon intercept is smaller than the one for the
BFKL Pomeron.

From an experimental point of view the Odderon remains a
mystery. Recent data on the differential elastic $pp$ scattering show
that one needs the Odderon to describe the cross sections in the dip
region \cite{dip-pp}. On the other hand, the perturbative QCD
prediction in the Born approximation for the cross section of
diffractive photoproduction of $\eta_c-$meson is rather small
\cite{eta}. The inclusion of evolution based on the BKP equation
increases the cross section by one order of magnitude
\cite{BBCV}. Because of that suggestions were raised to look for
Odderon effects in the studies of the charge asymmetries in the two
pion diffractive production \cite{HPST} and $\eta_c$ electroproduction
in the triple Regge region \cite{BBV}. A comprehensive survey of
searches for Odderon effects as well as its present day theoretical
status is given in the recent review \cite{Ewerz}.

The original derivation of the BFKL equation was based on the LLA
resummation of conventional Feynman diagrams in the Regge kinematics
typical for high energy and diffractive processes. Another approach to
the description of high energy processes was proposed by Mueller, who
constructed the dipole model \cite{dip} based on the light cone
perturbation theory. In the dipole model, the Pomeron, or in other
words the BFKL evolution emerges as an evolution of the wave function
of a projectile particle as viewed in the reference frame of a target
particle.  In Ref. \cite{dipBFKL} Navelet and Wallon have shown the
equivalence of the color dipole model and the BFKL Pomeron. Since the
dipole model is actually a very popular and efficient tool for the
description of high energy and diffractive processes and a starting
point for inclusion of saturation effects \cite{glr,yuri} it appears
natural to ask whether the Odderon can also be incorporated in the
dipole model similar to how the BKP equation was derived within the
same approach that had earlier led to the BFKL equation. Finding the
answer to this question is the main motivation for the present work.

The paper is structured as follows. In Sect. 2 we demonstrate that the
Odderon evolution is described in the dipole model by the same
evolution equation (\ref{oddeq}) as was used in \cite{dip} to describe
the BFKL evolution. The only difference is that to project out the
Odderon evolution from this equation one needs to impose $C$-odd
initial conditions (\ref{inc2}). The solution of \eq{oddeq} is given
by \eq{gensol} using the eigenfunctions (\ref{enn}) and the
eigenvalues (\ref{eig}) for odd $n$. The leading high energy Odderon
intercept is equal to $1$ as shown in \eq{oddint}. In Sect. 3 we show
that our solution (\ref{3gsol}) is equivalent to the BLV Odderon
solution \cite{BLV}. Thus, for a given three-gluon initial condition,
the dipole BFKL equation is equivalent to the BKP equation for three
$t$-channel gluons. In Sect. 4 we write down the equation
(\ref{oddnl}) describing the Odderon evolution including the
saturation effects. Saturation is likely to make the Odderon solution
a decreasing function of energy. We conclude in Sect. 5 by discussing
what we have and have not proved.

\section{Evolution Equation For Perturbative QCD Odderon in the Dipole Model}

In this Section we show that in Mueller's dipole model the odderon
evolution equation is given by the dipole model equivalent of the BFKL
equation taken with a $C$-odd initial condition. Solutions of this
equation for the odderon exchange amplitude are given by functions
$E^{n, \nu}$ with odd $n$. Corresponding eigenvalues are given by the
BFKL equation eigenvalues $\chi (n, \nu)$ with odd $n$, such that the
leading high energy intercept is given by $\alpha_{odd} - 1 = \frac{2
\, \as \, N_c}{\pi} \, \chi (n=1 ,\nu=0) = 0$.

\subsection{One Step of Small-$x$ Evolution}

Let us consider diffractive scattering of a $q\bar q$ dipole on some
target, which could be another dipole for the case of onium--onium
scattering or a proton for the case of Deep Inelastic Scattering
(DIS).  For the scattering amplitude of a $q\bar q$ dipole on a target
the operation of charge conjugation corresponds to interchanging the
quark and the anti-quark lines. Consider a dipole consisting of a
quark at transverse coordinate ${\un x}_0$ and an anti-quark at
transverse coordinate ${\un x}_1$ carrying the light cone momentum
fractions $z$ and $1-z$ of the total light cone momentum of the $q\bar
q$ system. Operation of charge conjugation corresponds to replacing
\be\label{C} 
C: \hspace*{1cm} {\un x}_0 \, \leftrightarrow \, {\un x}_1,
\hspace*{1cm} z \, \leftrightarrow \, 1-z.
\ee
Odderon exchange, by definition, corresponds to diffractive amplitudes
which are anti-symmetric under the operation shown in
\eq{C}:
\be\label{Codd}
{\cal O} ({\un x}_0, {\un x}_1; z, 1-z) \, = \, - \, {\cal O} ({\un
x}_1, {\un x}_0; 1-z, z),
\ee
where ${\cal O} ({\un x}_0, {\un x}_1; z, 1-z)$ is the forward
amplitude of a dipole $01$ on some target. Below we will use \eq{Codd}
as a definition of the odderon--mediated amplitude.

\begin{figure}[h]
\begin{center}
\epsfig{file=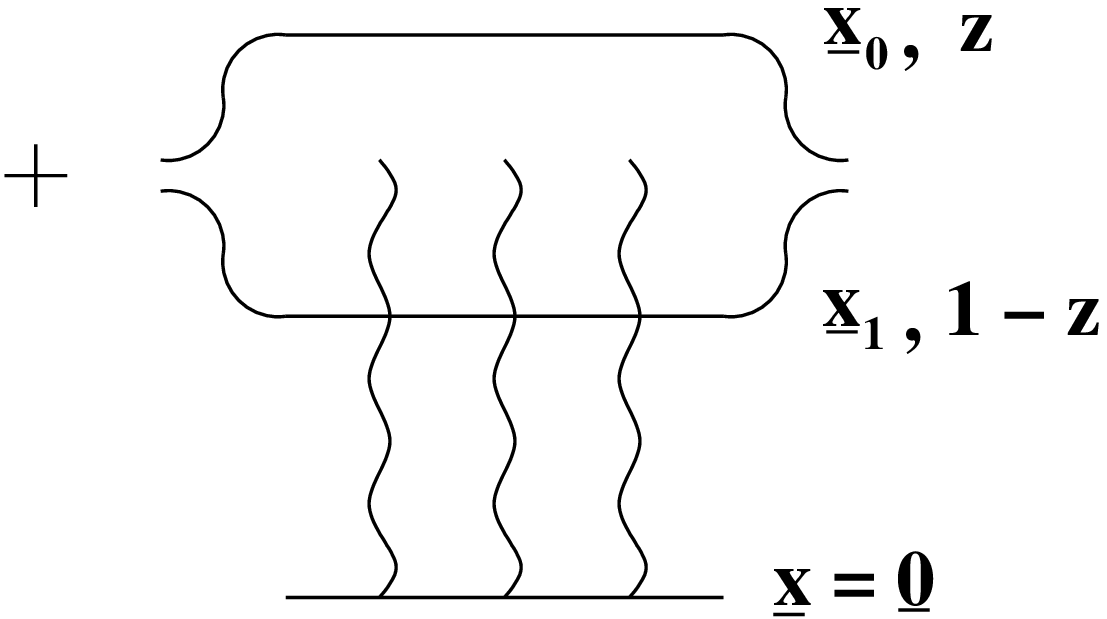, width=7cm}
\end{center}
\caption{Lowest order odderon exchange diagram which will be used as 
initial condition for small-$x$ evolution. The gluon lines connect to
both the quark and the anti-quark in the onium above.}
\label{init}
\end{figure}

Our goal now is to start with some $C$-odd initial condition for
onium--target scattering process and try to include one step of the
dipole evolution \cite{dip,CM} in it. The simplest initial condition
is given by the color-singlet three-gluon exchange diagram shown in
\fig{init} corresponding to the lowest order odderon exchange. Here,
for simplicity, we take the target to be a single quark line at the
bottom of \fig{init} located at transverse coordinate ${\un x} = {\un
0}$. In \fig{init} the disconnected $t$-channel gluon lines imply
summation over all possible connections of these lines to both the
quark and the anti-quark in the onium state at the top of the figure,
as shown in \fig{impact}. When convoluted with virtual photon's wave
function on the left hand side and with a meson's wave function on the
right hand side, the diagram in \fig{init} would contribute to
production of pions, $\eta_c$, etc., in DIS (see
e.g. \cite{BLV,BBCV}).

The three-gluon exchange diagram in \fig{init} gives the imaginary
part of the forward odderon-mediated amplitude ($x_i = |{\un x}_i|$)
\be\label{inc}
{\cal O}_0 ({\un x}_0, {\un x}_1) \, = \, c_0 \, \as^3 \, \ln^3
\frac{x_0}{x_1}
\ee
with
\be\label{c0}
c_0 = \frac{(N_c^2 - 4) \, (N_c^2 - 1)}{4 \, N_c^3}.
\ee
The color structure of the color-singlet three gluon exchange is given
by $d^{abc}$ for the odderon amplitude. The calculation leading to
\eq{inc} can be performed in covariant gauge, as well as in the $A^+ =
0$ light cone gauge which we will be using throughout the paper. (The
onium is moving in the light cone $+$ direction.) One can easily see
that the amplitude in \eq{inc} satisfies the condition of \eq{Codd}
and is, therefore, $C$-odd. One can also verify that the integral of
the amplitude (\ref{inc}) over the impact parameter ${\un b} = ({\un
x}_0 + {\un x}_1)/2$ is zero, in agreement with the fact that the
odderon exchange is zero at momentum transfer $t=0$.

\begin{figure}[h]
\begin{center}
\epsfig{file=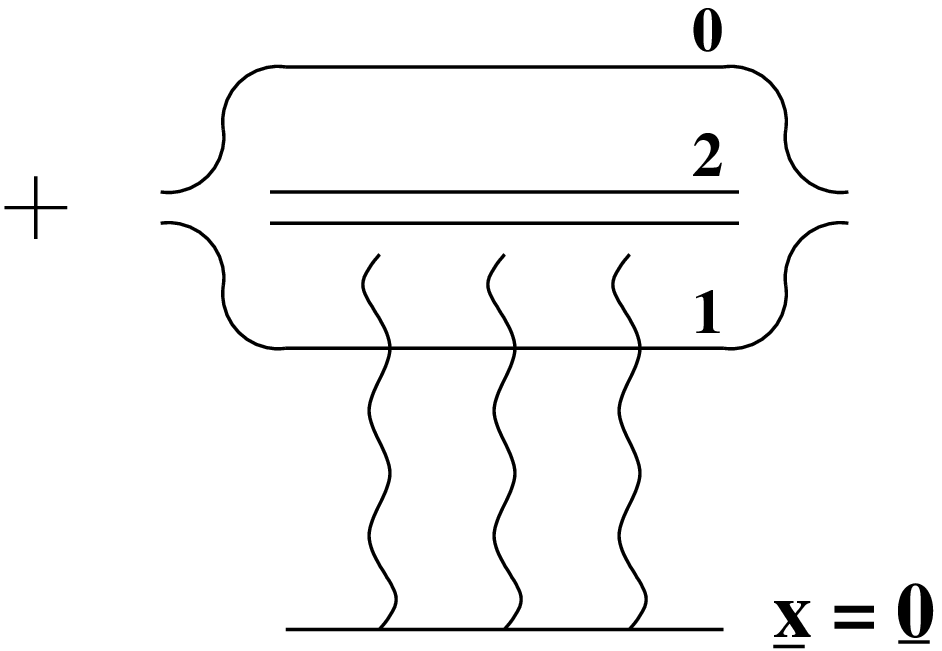, width=6cm}
\end{center}
\caption{One step of dipole evolution with the three-gluon exchange 
initial conditions. }
\label{one}
\end{figure}

One step of the dipole evolution \cite{dip} corresponds to emitting a
single $s$-channel gluon in the onium wave function. In the
large-$N_c$ limit in which the dipole model \cite{dip} is constructed,
the gluon line is represented by a double quark line, as shown in
\fig{one}. The diagram in \fig{one} represents a real part of the dipole 
kernel, where the emitted gluon is present at the time of scattering
on the target. The $s$-channel gluon can be emitted by either quark or
anti-quark to the left and to the right of the interaction, which is implied
by the disconnected gluon (double) line. The original dipole $01$ is
split into two (color-singlet) dipoles $02$ and $21$. Three exchanged
gluons can only couple all together to either one of the dipoles $02$
and $21$. \fig{one} shows the interaction with the dipole $21$.  The
only other possible interaction diagram involves one of the gluons
coupling to one of the dipoles with the other two gluons coupling to
another dipole. That diagram is zero since a color-neutral dipole can
not interact with the target by a single gluon exchange.

Employing the dipole evolution kernel from \cite{dip} we obtain the
following expression for the first step of dipole evolution with the
initial condition (\ref{inc}) (${\un x}_{ij} = {\un x}_i - {\un x}_j$)
\be\label{step1}
\frac{\as \, N_c}{2 \, \pi^2} \, \int^{\mbox{min} \{z,1-z\}}_{z_{init}} 
\frac{d z_2}{z_2} \, \int d^2 x_2 \, \frac{x_{01}^2}{x_{02}^2 \,
x_{12}^2} \, \left[ {\cal O}_0 ({\un x}_0, {\un x}_2) + {\cal O}_0
({\un x}_2, {\un x}_1) - {\cal O}_0 ({\un x}_0, {\un x}_1) \right]
\ee
where the first and the second terms correspond to real emissions with
the second term pictured in \fig{one}, while the last term in
\eq{step1} corresponds to the virtual correction \cite{dip}. In \eq{step1} 
$z_2$ is the fraction of onium's longitudinal (``plus'') momentum
carried by the gluon $2$, which, in the leading logarithmic
approximation, is integrated from some initial value $z_{init}$ up to the
smaller one of the light cone momentum fractions carried by the
original quark and anti-quark \cite{dip}. Substituting ${\cal O}_0$
from \eq{inc} into \eq{step1} yields
\be\label{step1'}
3 \, \frac{\as \, N_c}{2 \, \pi^2} \, c_0 \, \as^3 \, \ln
\frac{x_0}{x_1} \ \ln \left( \frac{\mbox{min} \{z,1-z\}}{z_{init}} \right) \, 
\int d^2 x_2 \, \frac{x_{01}^2}{x_{02}^2 \,
x_{12}^2} \, \ln \frac{x_0}{x_2} \, \ln \frac{x_1}{x_2}.
\ee
The amplitude in \eq{step1'} is non-zero. It changes sign under the
transformation of \eq{C} and is thus $C$-odd. We conclude that $C$-odd
initial conditions leave a dipole amplitude $C$-odd even after one
step of small-$x$ dipole evolution. The resulting amplitude is
non-zero, which means that the projection of the dipole evolution on
the $C$-odd channel is non-trivial.

\subsection{Evolution Equation}

It is now straightforward to write down an evolution equation for the
odderon in the dipole model. We have shown above that in the dipole
model a single step of the small-$x$ evolution for the odderon is the
same as for the BFKL pomeron \cite{dip}. Subsequent steps of the
dipole evolution would generate more color dipoles in the onium wave
function. Similar to one step of evolution shown in \fig{one}, there
is only one way of connecting the three-gluon initial conditions to
the fully evolved dipole wave function having many dipoles in it: only
the diagrams with all three gluons interacting with the same dipole
survive, since a dipole can not interact via a single gluon
exchange. Therefore, the picture of the odderon evolution is
simple. The onium wave function develops the usual dipole cascade
described by the leading logarithmic dipole evolution
\cite{dip}. In the linear evolution approximation, one of the dipoles 
generated by the evolution interacts with the target. In the case of
the pomeron exchange, the lowest order interaction is given by a
two-gluon exchange \cite{dip}. For the case of the odderon considered
here, the interaction should be $C$-odd and at the lowest order is
given by the three-gluon exchange amplitude (\ref{inc}).

For the odderon evolution equation we thus write \cite{dip}
\be\label{oddeq}
\frac{\partial}{\partial Y} {\cal O} ({\un x}_0, {\un x}_1, Y) \, = \, 
\frac{\as \, N_c}{2 \, \pi^2} \, \int d^2 x_2 \, 
\frac{x_{01}^2}{x_{02}^2 \, x_{12}^2} \, \left[ {\cal O} ({\un x}_0, 
{\un x}_2, Y) + {\cal O} ({\un x}_2, {\un
x}_1, Y) - {\cal O} ({\un x}_0, {\un x}_1, Y) \right]
\ee
where we changed the arguments of the dipole amplitude ${\cal O} ({\un
x}_{i}, {\un x}_j; z_i, z_j)$ to ${\cal O} ({\un x}_{i}, {\un x}_j,
Y)$ such that $Y = \ln (\mbox{min} \{z_i, z_j \} / z_{init})$ is the
rapidity variable. In the leading logarithmic approximation for gluon
evolution the dependence of the dipole amplitude on the light cone
momentum fractions of the quark $z_i$ and the antiquark $z_j$ comes in
only through the quantity $\mbox{min} \{z_i, z_j \}$ \cite{dip}, which
allowed us to reduce the number of arguments in ${\cal O}$.

\begin{figure}[h]
\begin{center}
\epsfig{file=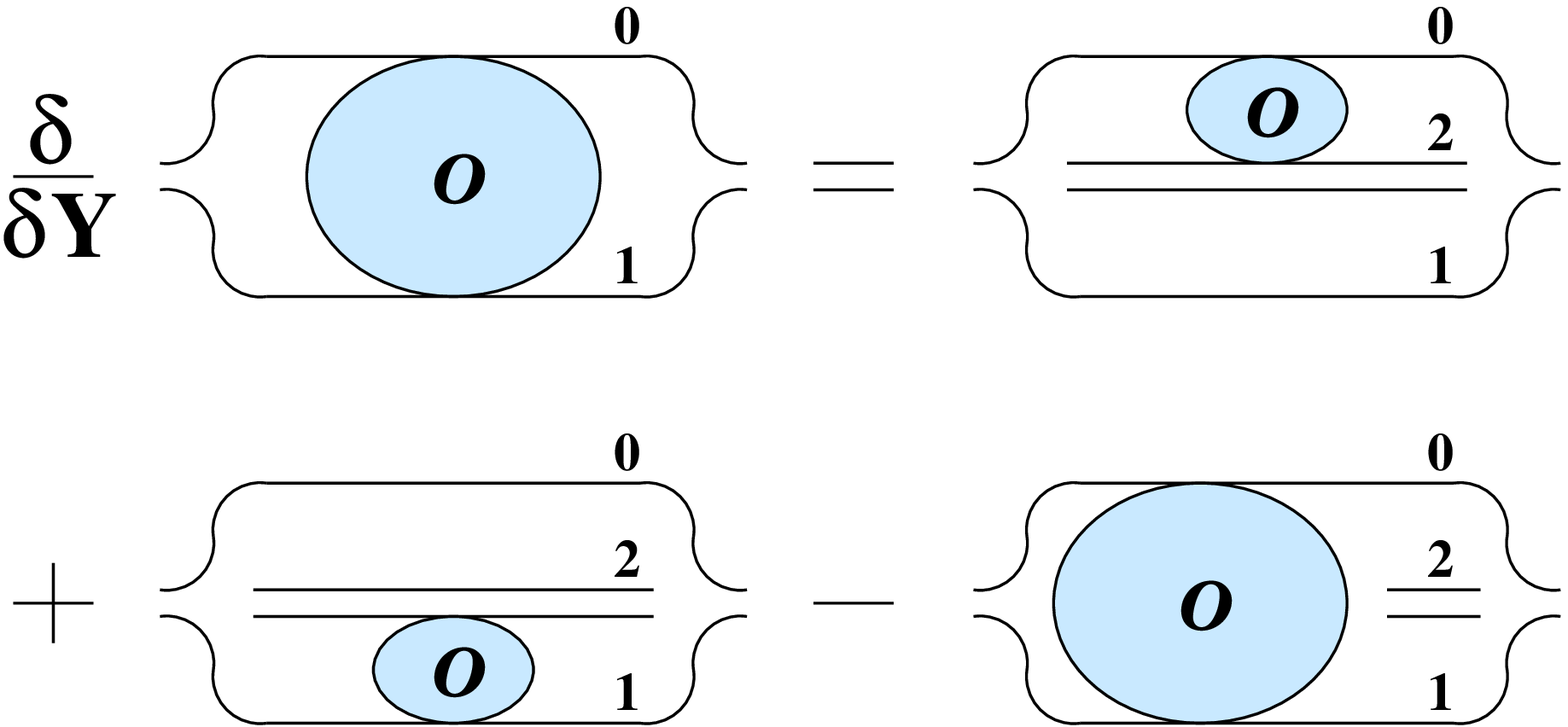, width=10cm}
\end{center}
\caption{Dipole evolution equation for the odderon exchange amplitude.}
\label{lineq}
\end{figure}

The initial condition for the \eq{oddeq} is given by 
\be\label{inc2}
{\cal O} ({\un x}_0, {\un x}_1, Y=0) \, = \, {\cal O}_0 ({\un x}_0,
{\un x}_1) \, = \, c_0 \, \as^3 \, \ln^3 \frac{x_0}{x_1}.
\ee
Eqs. (\ref{oddeq}) and (\ref{inc2}) describe the perturbative QCD
odderon in the color dipole model. \eq{oddeq} is identical to the
dipole model equivalent of the BFKL equation found in \cite{dip}.

\eq{oddeq} is illustrated in \fig{lineq}. The blob on the left hand side 
of \fig{lineq} represents the dipole evolution and the interaction
with the target leading to the odderon exchange amplitude for the
dipole $01$. After one step of the evolution a soft gluon is
emitted. This is shown in the three terms on the right hand side of
\fig{lineq}. They all correspond to the three terms on the right of
\eq{oddeq}. The first two terms correspond to real emissions, where
the subsequent evolution continues either in dipole $02$ or in dipole
$21$. The last term represents virtual corrections, where the gluon
can be emitted and absorbed on either side of the blob.

The rapidity variable does not change under $C$-parity transformation
of \eq{C}. Therefore, in the leading logarithmic approximation
considered here, charge conjugation only interchanges the transverse
coordinates of the quark and the anti-quark. Thus one can explicitly
see in \eq{oddeq} that substituting a $C$-odd amplitude into its right
hand side yields a $C$-odd expression, as we have seen in the previous
section for the case of three-gluon exchange amplitude. Evolution of
\eq{oddeq} preserves $C$-parity of the amplitude. When the initial
condition for \eq{oddeq} is given by some $C$-even amplitude, such as
the two-gluon exchange, it projects out the $C$-even solution of this
equation, which corresponds to the BFKL pomeron \cite{dip}. $C$-odd
initial condition projects out a $C$-odd solution, which corresponds
to the perturbative odderon exchange.

A comment is in order here. In \cite{CM} a dipole model equation was
constructed which was equivalent to the BKP equation \cite{BJKP} for
a four-reggeon system. The resulting equation (48) of \cite{CM} is
somewhat more complicated than the dipole BFKL equation
(\ref{oddeq}). One may wonder why there appears to be no such increase
of complication in the dipole model in going from a two-reggeon
(pomeron) system to the three reggeon (odderon) system. The reason for
that is clear. To describe the four-reggeon state in the dipole model
the authors of \cite{CM} had to make sure that this state did not mix
with the single pomeron exchange, double pomeron exchange and the
odderon. Imposing a four-gluon exchange initial conditions in
\cite{CM} made sure that the state had a positive parity $C=+1$ and
did not mix with the odderon having $C=-1$. To avoid mixing with the
two-pomeron state the authors of \cite{CM} had to make sure that the
state had a single-cylinder color topology. Finally, to avoid mixing
of this single-cylinder state with the single pomeron the authors of
\cite{CM} constructed a coupling of four gluons to the dipole wave
function which preserved the single-cylinder topology without
connecting all four gluons to the same dipole. That required
introduction of color quadrupoles and resulted in a more complicated
kernel for the integral equation. Now the three-reggeon state
considered here is much simpler: the only potential problem is for it
to mix with the pomeron, but that can never happen due to different
$C$-parities. Therefore we do not have to invent any complicated
color-quadrupole coupling of the three-gluon state to the onium wave
function. While it exists, it is $N_c$-suppressed compared to the
dipole coupling considered above.

\subsection{Odderon Solution}

The solution of \eq{oddeq} is easy to construct. We first note that
the eigenfunctions of the Casimir operators of conformal algebra
\be\label{enn}
E^{n, \nu} (\rho_0, \rho_1) \, = \, \left( \frac{\rho_{01}}{\rho_0 \,
\rho_1} \right)^{\frac{1+n}{2} + i \nu} \ \left(
\frac{\rho_{01}^*}{\rho_0^* \, \rho_1^*} \right)^{\frac{1-n}{2} + i
\nu}
\ee
are also the eigenfunctions of the dipole kernel of \eq{lineq} with
the eigenvalues
\be\label{eig}
2 \, \bas \, \chi (n, \nu),
\ee 
where 
\be\label{chi}
\chi (n, \nu) \, = \, \psi (1) - \frac{1}{2} \, \psi \left( 
\frac{1+|n|}{2} + i \nu \right)  - \frac{1}{2} \, \psi \left( 
\frac{1+|n|}{2} - i \nu \right),  
\ee
\be
\bas \, = \, \frac{\as \, N_c}{\pi},
\ee
$\rho_{ij} = \rho_i - \rho_j$ and the asterisk in \eq{enn} denotes
complex conjugation. \eq{eig} follows from \eq{goal} which is derived
in Appendix A.

\begin{figure}[h]
\begin{center}
\epsfig{file=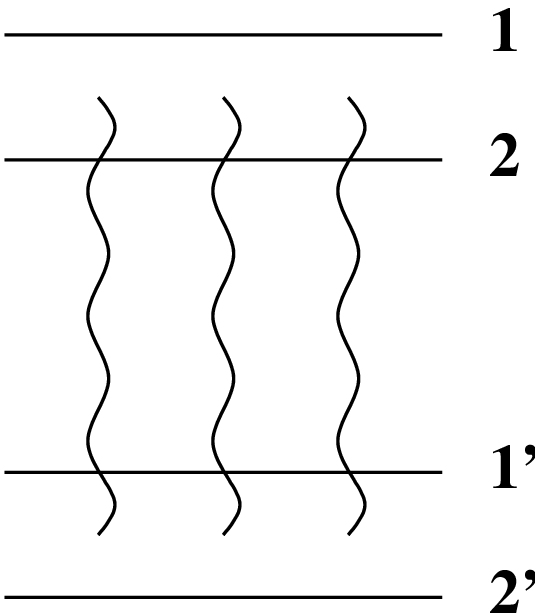, width=3cm}
\end{center}
\caption{Onium-onium scattering.}
\label{oo}
\end{figure}

To construct the solution of \eq{oddeq} with a three-gluon exchange
initial conditions let us consider onium-onium scattering, i.e. take
the target of the above discussion to be a $q\bar q$ pair as depicted
in \fig{oo}. We will work in the frame where the target onium $1'2'$ is
at rest, so that all the small-$x$ evolution will be included in the wave
function of the incoming onium $12$.

A general solution of \eq{oddeq} is easy to write down once we know
that functions $E^{n, \nu}$ from \eq{enn} are the eigenfunctions of
its kernel. Since
\be
E^{n, \nu} (\rho_{1}, \rho_{2}) \, = \, (-1)^n \, E^{n, \nu}
(\rho_{2}, \rho_{1})
\ee
only the functions $E^{n, \nu} (\rho_{1}, \rho_{2})$ with odd $n$
satisfy the condition (\ref{Codd}). The most general $C$-odd solution
reads in the complex notation
\be\label{gensol}
{\cal O} (\rho_1, \rho_2 ; \rho_{1'}, \rho_{2'}; Y) \, = \,
\sum_{\mbox{odd n}} \, \int_{-\infty}^\infty d \nu \, \int d^2
\rho_0 \, e^{2 \, \bas \, \chi(n, \nu) \, Y} \, C_{n, \nu}  \, 
E^{n, \nu} (\rho_{10}, \rho_{20}) \, 
E^{n, \nu \, *} (\rho_{1'0}, \rho_{2'0}),
\ee
where the coefficients $C_{n, \nu}$ have to be determined from the
initial conditions. The equivalent of \eq{inc2} for the dipole-dipole
scattering of \fig{oo} reads in complex notation
\be\label{incompl}
{\cal O} (\rho_1, \rho_2 ; \rho_{1'}, \rho_{2'}; Y=0) \, = \, c_0 \,
\as^3 \, \ln^3 \bigg| \frac{\rho_{11'} \, \rho_{22'}}{\rho_{12'} \, 
\rho_{1'2}} \bigg|.
\ee
To determine the coefficients $C_{n, \nu}$ we rewrite the amplitude
(\ref{incompl}) as (see Appendix B)
\ben
{\cal O} (\rho_1, \rho_2 ; \rho_{1'}, \rho_{2'}; Y=0) \, = \, c_0 \,
\as^3 \, \frac{6}{\pi^2} \, \sum_{\mbox{odd n}} \, \int_{-\infty}^\infty 
d \nu \, \int d^2
\rho_0 \, \frac{\nu^2 + \frac{n^2}{4}}{\left[ \nu^2 + \frac{(n+1)^2}{4} 
\right] \, \left[ \nu^2 + \frac{(n-1)^2}{4} 
\right]}
\een
\be\label{decom}
\times \, \chi (n, \nu) \, E^{n, \nu} (\rho_{10}, \rho_{20}) \, 
E^{n, \nu \, *} (\rho_{1'0}, \rho_{2'0})
\ee
which yields
\be\label{coef}
C_{n, \nu} \, = \,  c_0 \, \as^3 \, \frac{6}{\pi^2} \, 
\frac{\nu^2 + \frac{n^2}{4}}{\left[ \nu^2 + \frac{(n+1)^2}{4} 
\right] \, \left[ \nu^2 + \frac{(n-1)^2}{4} \right]} \, \chi (n, \nu).
\ee
Using \eq{coef} in \eq{gensol} we find the solution of \eq{oddeq} with
the initial condition (\ref{incompl})
\ben
{\cal O} (\rho_1, \rho_2 ; \rho_{1'}, \rho_{2'}; Y) \, = \, c_0 \,
\as^3 \, \frac{6}{\pi^2} \, \sum_{\mbox{odd n}} \, \int_{-\infty}^\infty 
d \nu \, \int d^2
\rho_0 \, \frac{\nu^2 + \frac{n^2}{4}}{\left[ \nu^2 + \frac{(n+1)^2}{4} 
\right] \, \left[ \nu^2 + \frac{(n-1)^2}{4} 
\right]}
\een
\be\label{3gsol}
\times \, e^{2 \, \bas \, \chi(n, \nu) \, Y} \, \chi (n, \nu) \,
E^{n, \nu} (\rho_{10}, \rho_{20}) \, 
E^{n, \nu \, *} (\rho_{1'0}, \rho_{2'0}).
\ee
Therefore, we conclude that the eigenfunctions and eigenvalues of the
dipole amplitude with the odderon exchange are given by
Eqs. (\ref{enn}) and (\ref{eig}) correspondingly, with only odd values of
$n$. Analyzing \eq{3gsol} one can easily see that the leading high
energy intercept of the odderon amplitude is given by
\be\label{oddint}
\alpha_{odd} - 1 \, = \, 2 \, \bas \, \chi(n=1, \nu =0) \, = \, 0
\ee
in agreement with the results of Bartels, Lipatov and Vacca (BLV)
\cite{BLV}.

\section{Connection to Traditional Approaches}

\begin{figure}
\begin{center}
\epsfig{file=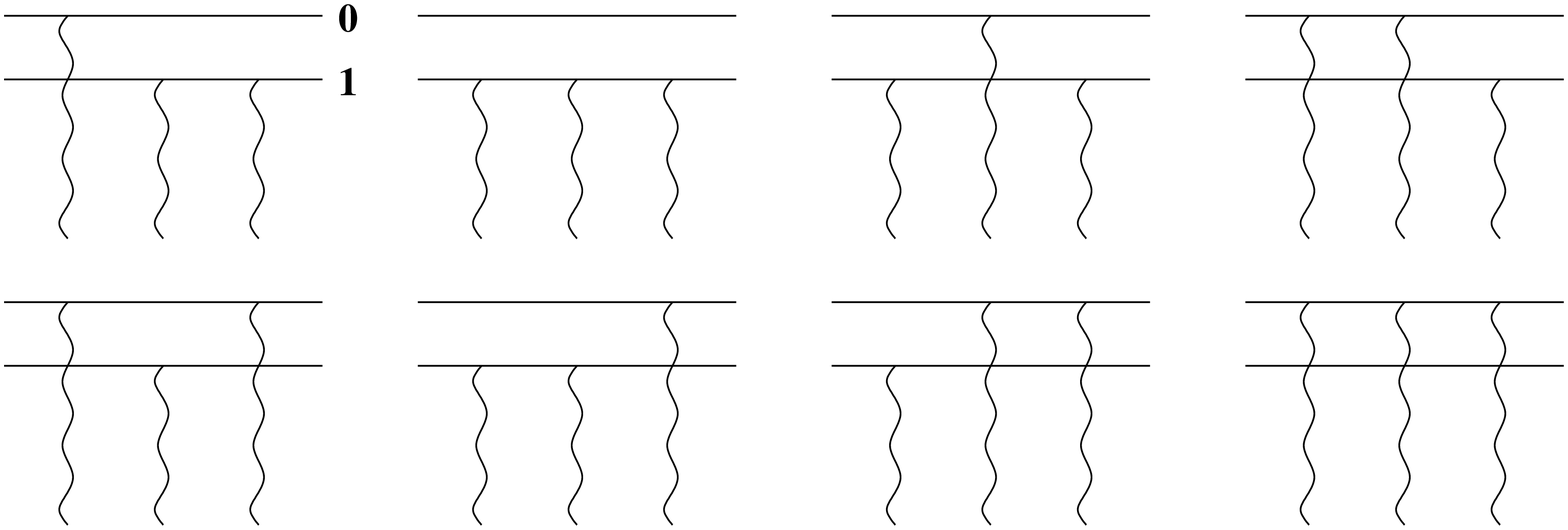, width=13cm}
\end{center}
\caption{Dipole's impact factor.}
\label{impact}
\end{figure}

To make a connection between the dipole evolution equation
(\ref{oddeq}) and the BKP equation \cite{BJKP} for the three-reggeon
state we will now show that the eigenfunctions from \eq{enn} with odd
$n$ correspond to the odderon solution found by Bartels, Lipatov and
Vacca in \cite{BLV}. To find the dipole scattering amplitude generated
by the odderon Green function we have to connect the external gluon
legs of the Green function to the dipole, as shown in
\fig{impact}. Instead of the full odderon Green function (see Eq. (5) in 
\cite{BBCV}), we take a single BLV eigenfunction in the form given by Eq. 
(2) of \cite{BBCV}
\be\label{blveig}
\Psi^{n,\nu} (k_1, k_2, k_3) \, = \, c (n,\nu) \, \sum_{(123)} \, 
\frac{({\un k}_1 + {\un k}_2)^2}{{\un k}_1^2 \, {\un k}_2^2} \, 
E^{n, \nu} ({\un k}_1 + {\un k}_2, {\un k}_3), \hspace*{1cm}
\mbox{odd} \ n,
\ee
and connect in to the dipole $x_{01}$ in all possible ways as depicted
in \fig{impact}. $c (n,\nu)$ in \eq{blveig} is the overall
normalization coefficient which is not important and is given in
\cite{BBCV}. Summation over all permutations of $1$, $2$ and $3$ is 
implied in \eq{blveig}. The dipole amplitude would then be given by
\ben
{\cal O}^{n, \nu} ({\un x}_0, {\un x}_1, Y) \, = \,  
\int \left[ \prod_{i=1}^3 \,  d^2 k_i \,
(e^{- i {\un k}_i \cdot {\un x}_0} - e^{- i {\un k}_i \cdot {\un
x}_1}) \right] \, \Psi^{n,\nu} (k_1, k_2, k_3) 
\een
\be\label{dip1}
\, = \, 3 \,  c (n,\nu) \, 
\int \left[ \prod_{i=1}^3 \,  d^2 k_i \,
(e^{- i {\un k}_i \cdot {\un x}_0} - e^{- i {\un k}_i \cdot {\un
x}_1}) \right] \, 
\frac{({\un k}_1 + {\un k}_2)^2}{{\un k}_1^2 \, {\un k}_2^2} \, 
E^{n, \nu} ({\un k}_1 + {\un k}_2, {\un k}_3). 
\ee
Substituting 
\be
E^{n, \nu} ({\un k}_1 + {\un k}_2, {\un k}_3) \, = \, \int \frac{d^2
r_1 \, d^2 r_2}{(2 \pi)^4} \, e^{i ({\un k}_1 + {\un k}_2) \cdot {\un
r}_1 + i {\un k}_3 \cdot {\un r}_2} \, E^{n, \nu} (r_1, r_2)
\ee
into \eq{dip1} and integrating over $k_i$'s yields 
\ben
{\cal O}^{n, \nu} ({\un x}_0, {\un x}_1, Y) \, = \, 6 \, c (n,\nu) \,
\int d^2 r_1 \, d^2 r_2 \, E^{n, \nu} (r_1, r_2) \,
\left\{
\frac{x_{01}^2}{|{\un x}_1 - {\un r}_1|^2 \, |{\un x}_0 - {\un r}_1|^2} \,
 \right.
\een
\be\label{dip2}
\times \, \left[ \delta ({\un x}_1 - {\un r}_2 ) - 
\delta ({\un x}_0 - {\un r}_2 ) \right]
- \left. 2 \pi \, \ln \frac{|{\un x}_1 - {\un r}_1|}{|{\un x}_0 -
 {\un r}_1|} \left[ \delta ({\un x}_0 - {\un r}_1 ) \, \delta ({\un
 x}_1 - {\un r}_2 ) + \delta ({\un x}_0 - {\un r}_2 ) \, \delta ({\un
 x}_1 - {\un r}_1 ) \right] \right\}.
\ee
Regulating $|{\un x}_1 - {\un r}_1| = 0$ and $|{\un x}_0 - {\un r}_1|
= 0$ singularities in the same way in the first and the second terms
on the right hand side of \eq{dip2} and integrating \eq{dip2} over
$r_2$ we rewrite it as
\ben
{\cal O}^{n, \nu} ({\un x}_0, {\un x}_1, Y) \, = \, 6 \, c (n,\nu) \,
\int d^2 r_1 \, \frac{x_{01}^2}{|{\un x}_1 - {\un r}_1|^2 \, |{\un
x}_0 - {\un r}_1|^2} \, \left[ E^{n, \nu} (x_0, r_1) + E^{n, \nu}
(r_1, x_1) \right.
\een
\be\label{dipeig}
\left. - E^{n, \nu} (x_0, x_1) \right]. 
\ee
Using \eq{goal} which is proved in Appendix A we rewrite \eq{dipeig}
as
\be\label{conn}
{\cal O}^{n, \nu} ({\un x}_0, {\un x}_1, Y) \, = \, 24 \,
c (n,\nu) \, \pi \, \chi(n, \nu) \, E^{n, \nu} (x_0, x_1).
\ee
\eq{conn} shows that momentum space BLV eigenfunctions in the 
$t$-channel Green function translate into the functions $E^{n, \nu}$
for the dipole scattering amplitude in transverse coordinate space up
to an overall normalization factor. Therefore our odderon solution
(\ref{gensol}) is equivalent to the BLV solution
\cite{BLV}. As one can explicitly check, taking the odderon Green 
function from Eq. (5) of \cite{BBCV}, connecting it to the two
colliding onia $12$ and $1'2'$ as shown at the lowest order in
\fig{oo}, and Fourier-transforming into transverse coordinate space 
as was done in arriving at \eq{conn} would yield us the solution in
\eq{3gsol}. Starting with the same three-gluon exchange initial 
condition (\ref{decom}) our evolution equation (\ref{oddeq}) and the
BKP equation \cite{BJKP} would give us the same answer for the
amplitude and are equivalent in the case of dipole scattering on a
target.

Since the most general $C$-odd solution of \eq{oddeq} is known and is
given by \eq{gensol}, we have proven that for the case of a dipole
scattering on a target there could be no other odderon solution except
for the one found above, or, equivalently, except for BLV solution
\cite{BLV}. Any $C$-odd dipole amplitude satisfying the condition of 
\eq{Codd} and being a solution of \eq{oddeq} can always be decomposed in 
a series (\ref{gensol}). Therefore, for the scattering of a dipole on
any target the resulting cross section can not grow with center of
mass energy $s$ faster than $s^0$ as follows from the intercept of
\eq{oddint}. (The saddle point evaluation of the $\nu$ integral around 
$\nu =0$ of the $n=1$ term in \eq{3gsol} would give an amplitude which
would grow slowly with energy, though the growths would be slower than
any power of energy.)

\section{Including Saturation Effects}

Consider DIS on a proton or a nucleus at high enough energy so that
parton saturation effects \cite{glr} are becoming important. In
\cite{yuri,bal} (see also \cite{JKLW,FILM,Braun1}) a non-linear evolution 
equation has been derived which resums all multiple pomeron exchanges
in the $C$-even forward scattering amplitude of a $q\bar q$ dipole on
the target. The equation is easier to visualize for a nuclear target
with atomic number $A$. As was discussed in \cite{yuri}, in the frame
where the nucleus is at rest all the small-$x$ dipole evolution takes
place in the incoming $q\bar q$ dipole wave function. The evolution,
taken in the leading logarithmic approximation, resums powers of $\as
Y$. The color dipoles generated by evolution rescatter on the nuclear
target. The rescattering in \cite{yuri} was given by the
Glauber-Mueller formula \cite{Mueller1}, which corresponds to
resummation of the parameter $\as^2 A^{1/3}$ \cite{mv,kjklw}. This
parameter arises if we limit rescattering of a dipole on each nucleon
to two-gluon exchange only \cite{kjklw}.

To construct a similar equation for the $C$-odd (odderon) exchange
amplitude one has to allow one of the dipoles to interact with one
nucleon via a three-gluon exchange. In the high energy regime when
$\as Y \sim 1$ and $\as^2 A^{1/3} \sim 1$, a single three-gluon
exchange would bring in one power of $\as^3 A^{1/3} \sim \as$, so that
the corresponding odderon exchange amplitude would be of the order
${\cal O} \sim \as \ll 1$. Below we limit our analysis to a single
odderon exchange: multiple odderon exchanges would be suppressed by
extra powers of $\as$. In fact, the $C$-even amplitude including a
pomeron splitting into two odderons which interact with the target
would be of the order $\as^2$, which is of the same order as the
so-called pomeron loop diagrams. Resummation of these diagrams is a
very hard unsolved problem (see the discussion in the second reference
in \cite{yuri}) and is beyond the scope of this paper.

\begin{figure}[h]
\begin{center}
\epsfig{file=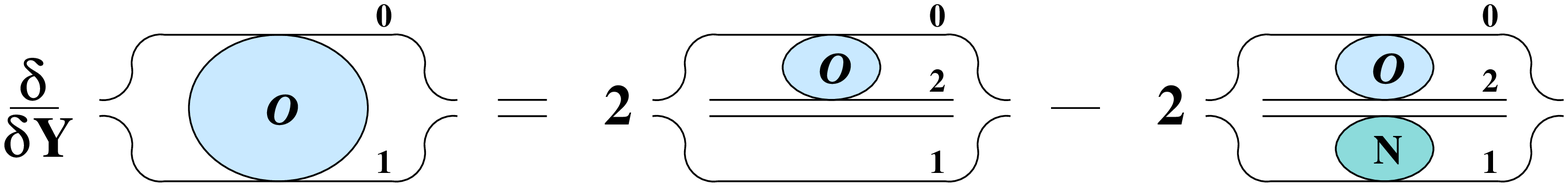, width=15cm}
\end{center}
\caption{Evolution equation for the odderon exchange amplitude ${\cal O}$ 
in the presence of gluon saturation, which comes in through the
$C$-even dipole (``pomeron'') amplitude $N$. }
\label{nl}
\end{figure}

The evolution equation for the odderon amplitude which includes
saturation effects is depicted in \fig{nl}. Similar to
\fig{lineq}, in one step of evolution a gluon is emitted splitting the
original dipole $01$ in two dipoles $02$ and $21$. The subsequent
odderon evolution can take place only in one of the dipoles. The first
term on the right hand side of \fig{nl} schematically represents all
three terms on the right hand side of \fig{lineq} and \eq{oddeq}. The
second term on the right hand side of \fig{nl} corresponds to the case
where the dipole without the odderon evolution develops nonlinear
$C$-even evolution in it. $N ({\un x}_0, {\un x}_1, Y)$ is the
$C$-even forward scattering amplitude of the dipole $01$ with rapidity
$Y$ on the target including all multiple rescatterings and pomeron
exchanges \cite{yuri}. $N ({\un x}_0, {\un x}_1, Y)$ is proportional
to the $T$-matrix of the interaction and is normalized in such a way
that it goes to $0$ when interactions are weak and goes to $1$ in the
black limit \cite{yuri}. The high energy evolution of $N ({\un x}_0,
{\un x}_1, Y)$ is described by the non-linear evolution equation
\cite{yuri,bal}. Here we assume that $N$ has been found from that
evolution equation and is known to us.

The graphical equation from \fig{nl} translates into  
\ben
\frac{\partial}{\partial Y} {\cal O} ({\un x}_0, {\un x}_1, Y) \, = \, 
\frac{\as \, N_c}{2 \, \pi^2} \, \int d^2 x_2 \, 
\frac{x_{01}^2}{x_{02}^2 \, x_{12}^2} \, \left[ {\cal O} ({\un x}_0, 
{\un x}_2, Y) + {\cal O} ({\un x}_2, {\un
x}_1, Y) - {\cal O} ({\un x}_0, {\un x}_1, Y) \right]
\een
\be\label{oddnl}
- \frac{\as \, N_c}{2 \, \pi^2} \, \int d^2 x_2 \, 
\frac{x_{01}^2}{x_{02}^2 \, x_{12}^2} \, \left[ {\cal O} ({\un x}_0, 
{\un x}_2, Y) \, N ({\un x}_2, {\un x}_1, Y) + N ({\un x}_0, 
{\un x}_2, Y) \, {\cal O} ({\un x}_2, {\un x}_1, Y) \right].
\ee
The initial condition for the amplitude ${\cal O}$ in \eq{oddnl} is
given by 
\be\label{inl1}
{\cal O} ({\un x}_0, {\un x}_1, Y=0) \, = \, c_0 \, \as^3 \, \rho \,
\int_{S_\perp} d^2 b \,  T({\un b}) \, \ln^3
\left( \frac{{\un B} - {\un b} + \frac{1}{2} \, {\un x}_{01}}{{\un B} - 
{\un b} - \frac{1}{2} \, {\un x}_{01}} \right) \, e^{- x_{01}^2 \, 
Q_s^2 (x_{01}, B) /4},
\ee
where $\rho$ is the atomic number density in the nucleus, $T(\un b)$
is the nuclear profile function ($T(\un b) = 2 \sqrt{R^2 - b^2}$ for a
spherical nucleus of radius $R$) and ${\un B} = ({\un x}_0 + {\un
x}_1)/2$. It includes a three gluon exchange interaction with a
nucleon (modeled by a single quark (cf. \eq{inc2})) averaged over
transverse positions of the nucleon in the nucleus. (The $b$-integral
in \eq{inl1} is not zero since the integration is limited by
transverse nuclear size $S_\perp$ and does not go out to infinity.) 
The exponential factor in Eq. (\ref{inl1}) accounts for multiple
$C$-even (two-gluon exchange) rescatterings of the dipole on other
nucleons in the nucleus
\cite{Mueller1,kjklw}, where
\be
Q_s^2 (x_{01}, B) = 4 \pi \, \as^2 \, \frac{C_F}{N_c} \, \rho \, T(\un B)
\, \ln \frac{1}{x_{01} \, \Lambda}
\ee
with $\Lambda$ some infrared cutoff.

\eq{oddnl} describes the odderon evolution including the effects of 
saturation in the leading logarithmic approximation. While a detailed
analysis of \eq{oddnl} is beyond the scope of this paper, we note that
gluon saturation introduces negative terms on the right hand side of
\eq{oddnl} (the last two terms in it). Since the leading high energy
intercept of \eq{oddeq} corresponding to the first three terms on the
right hand side of (\ref{oddnl}) is zero, the negative non-zero last
two terms on the right hand side of (\ref{oddnl}) are likely to make
the odderon amplitude a decreasing function of energy/rapidity. This
result may be relevant to the lack of observation of QCD odderon in
the high energy DIS and $pp$ data.

\section{Discussion}

In this paper we have shown that in the framework of the dipole model
both the Pomeron and the Odderon are described by the same evolution
equation (\ref{oddeq}). While the Pomeron evolution is projected out of
\eq{oddeq} by $C$-even initial conditions \cite{dip}, the Odderon evolution 
is projected out by the $C$-odd ones. Since the complete set of
solutions to the BFKL equation is known, we have constructed the most
general Odderon solution for the dipole amplitude given by
\eq{gensol}. We have thus showed that there could be no other Odderon 
solution for the case of a dipole scattering on any target. Our
solution (\ref{gensol}) is equivalent to the BLV solution \cite{BLV},
as shown in Sect. 3.

In the usual Feynman diagram language the dipole model picks out a
part of the Odderon Green function with at least two out of three
$t$-channel gluons connecting to the same (anti)-quark line. Therefore
at least two gluons have the same transverse coordinate. This happens
for the Odderon exchange in DIS. However, in the case of $pp$ or
$p\bar p$ scattering there exist Feynman diagrams with the three
$t$-channel gluons coupling to different quarks at different
transverse coordinates. These diagrams have no equivalent in the
dipole model. One can not reproduce the solution of \cite{JW} in our
approach since it vanishes when two gluons are taken at the same
transverse point. While the dipole model allowed us to find the most
general Odderon solution for the case of dipole scattering on any
target, it did not produce any upper bound on the intercept of the
Odderon exchanged in the scattering of two three-quark states ($pp$ or
$p\bar p$).

\section*{Acknowledgments} 

We want to thank Al Mueller for a number of critical discussions. We
are grateful to Ian Balitsky, Jochen Bartels, Misha Braun, Genya Levin
and Larry McLerran for encouraging us to study this
problem. L.Sz. thanks the Institute for Nuclear Theory and the Nuclear
Theory Group at the University of Washington for their hospitality and
the US Department of Energy for partial support during the beginning
of this work. He thanks also Oleg Teryaev for discussions and
hospitality at BLTP, JINR in Dubna, where this work was completed.

The work of Yu.K. was supported in part by the U.S. Department of
Energy under Grant No. DE-FG03-97ER41014. The work of L.Sz. was
supported by the French--Polish Scientific Agreement Polonium.

\renewcommand{\theequation}{A\arabic{equation}}
  \setcounter{equation}{0}  
  \section*{Appendix A: Eigenfunctions of the Dipole Kernel}
\label{eigenftn}

In this appendix we show that functions $E^{n, \nu} (x_0, x_1)$ are
eigenfunctions of the dipole kernel, such that
\be\label{goal}
\int d^2 x_2 \, \frac{x_{01}^2}{x_{02}^2 \,
x_{12}^2} \, \left[ E^{n, \nu} (x_0, x_2) + E^{n, \nu} (x_2, x_1) -
E^{n, \nu} (x_0, x_1) \right] \, = \, 4 \, \pi \, \chi(n, \nu) \,  
E^{n, \nu} (x_0, x_1)
\ee
with $\chi(n, \nu)$ given by \eq{chi}. Our derivation below will be
closely following the Appendix of \cite{Lipatov1} (see also
\cite{DF}). In the complex notation the integral (\ref{goal}) can be
written as
\be\label{II}
I (\rho_0, \rho_0^*, \rho_1, \rho_1^*) \, \equiv \, \int d^2 \rho_2 \,
\frac{|\rho_{01}|^2}{|\rho_{02}|^2 \, |\rho_{12}|^2} \, \left[ E^{n, \nu}
(\rho_0, \rho_2) + E^{n, \nu} (\rho_2, \rho_1) - E^{n, \nu} (\rho_0,
\rho_1) \right]
\ee
where the asterisk denotes complex conjugation. Performing the
inversion transformation $\rho_i \rightarrow 1/\rho_i$ yields
\be\label{inv}
I (1/\rho_0, 1/\rho_0^*, 1/\rho_1, 1/\rho_1^*) \, = \, \int d^2 \rho_2
\, \frac{|\rho_{01}|^2}{|\rho_{02}|^2 \, |\rho_{12}|^2} \, \left[ 
\rho_{20}^h \, \rho_{20}^{* \, \bar h} + \rho_{12}^h \, 
\rho_{12}^{* \, \bar h} - \rho_{10}^h \, \rho_{10}^{* \, \bar h} \right]
\ee
where $h = \frac{1+n}{2} + i \nu$ and ${\bar h} = \frac{1-n}{2} + i
\nu$. Defining a new variable
\be
R \, = \, \frac{1}{2} \, (\rho_{20} + \rho_{21})
\ee
we rewrite \eq{inv} as
\ben
I (1/\rho_0, 1/\rho_0^*, 1/\rho_1, 1/\rho_1^*) \, = \,  \int d^2 R \,
\frac{|\rho_{01}|^2}{|R - \frac{1}{2} \rho_{01}|^2 \, |R +  \frac{1}{2}
\rho_{01}|^2} \,  \Bigg[ |R - \frac{1}{2} \rho_{01}|^{1 + 2 i \nu}
\een
\be\label{rint}
\times \,
\left( \frac{R - \frac{1}{2} \rho_{01}}{R^* - \frac{1}{2} \rho_{01}^*} 
\right)^{n/2} + (-1)^n \, |R + \frac{1}{2} \rho_{01}|^{1 + 2 i \nu} 
\left( \frac{R + \frac{1}{2} \rho_{01}}{R^* + \frac{1}{2} \rho_{01}^*} 
\right)^{n/2} - \left. |\rho_{10}|^{1 + 2 i \nu} \left(
\frac{\rho_{10}}{\rho_{10}^*}
\right)^{n/2} \right].
\ee
Redefining the integration variable again such that $R = z \rho_{01}
/2$ we obtain
\ben
I (1/\rho_0, 1/\rho_0^*, 1/\rho_1, 1/\rho_1^*) \, = \, 4 \,  
|\rho_{10}|^{1 + 2 i \nu} \left( \frac{\rho_{10}}{\rho_{10}^*}
\right)^{n/2} \, \int d^2 z \, \frac{1}{(z^2 - 1) \, (z^{* \, 2} - 1)} \, 
\Bigg[ (-1)^n \, 2^{-1 - 2 i \nu} 
\een
\be\label{zint}
\times \, \left. |z-1|^{1 + 2 i \nu} \, \left( \frac{z-1}{z^* -1} 
\right)^{n/2} + 2^{-1 - 2 i \nu} \, |z+1|^{1 + 2 i \nu} \, 
\left( \frac{z+1}{z^* +1} \right)^{n/2} - 1 \right].
\ee
Undoing the inversion transformation $\rho_i \rightarrow 1/\rho_i$ in
\eq{zint} gives
\be\label{eigen}
I (\rho_0, \rho_0^*, \rho_1, \rho_1^*) \, = \, C (n, \nu) \, E^{n,
\nu} (\rho_0, \rho_1)
\ee
with
\ben
C (n, \nu) \, = \, 4 \, \int d^2 z \, \frac{1}{(z^2 - 1) \, (z^{* \,
2} - 1)} \, \left[ (-1)^n \, 2^{-1 - 2 i \nu} \, |z-1|^{1 + 2 i \nu} \, 
\left( \frac{z-1}{z^* -1} \right)^{n/2} \right.
\een
\be\label{c1}
+ \left. 2^{-1 - 2 i \nu} \, |z+1|^{1 + 2 i \nu} \, \left(
\frac{z+1}{z^* +1} \right)^{n/2} - 1 \right].
\ee
\eq{eigen} proves that functions $E^{n, \nu} (\rho_0, \rho_1)$ are, 
indeed, eigenfunctions of the dipole kernel. To find the eigenvalue $C
(n, \nu)$ we follow the steps outlined in the Appendix of
\cite{Lipatov1}. Rewriting $z=x + i y$ with real $x$ and $y$ we perform 
Wick rotation in the complex $y$-plane. The $y$-integral transforms
into an integral over $t = - i y$. Following \cite{Lipatov1} we define
$\alpha = x-t$ and $\beta = x+t$ obtaining
\ben
C (n, \nu) \, = \, 2 \, i \, \lim_{\sigma \rightarrow 0} \,
\int_{-\infty}^\infty d \alpha \, d \beta \, 
\frac{1}{[(1-\alpha) (1-\beta) + i \epsilon ]^{1-\sigma} \, 
[(1+\alpha) (1+\beta) + i \epsilon ]^{1-\sigma}}
\een
\ben
\times \, \left\{ 2^{-1 - 2 i \nu} \, (1-\alpha)^{|n|} \, 
[(1-\alpha) (1-\beta) 
+ i \epsilon ]^{\frac{1-|n|}{2} + i \, \nu} \right.
\een
\be\label{c2}
\left. +  2^{-1 - 2 i \nu} \, (1+\alpha)^{|n|} \, [(1+\alpha) (1+\beta) 
+ i \epsilon ]^{\frac{1-|n|}{2} + i \, \nu} - 1 \right\},
\ee
where we introduced a dimensional regulator $\sigma$ to be able to
integrate the terms separately. The $\beta$-contour in \eq{c2} can be
distorted to give $0$ unless $-1 < \alpha <1$. Noting that the first
and the second terms in the curly brackets of \eq{c2} are identical we
rewrite it as
\ben
C (n, \nu) \, = \, 2 \, i \, \lim_{\sigma \rightarrow 0} \,
\int_{-1}^1 d \alpha \, \frac{1}{(1-\alpha^2)^{1-\sigma}} \, 
\int_{-\infty}^\infty d \beta \, 
\frac{1}{(1-\beta + i \epsilon )^{1-\sigma} \, 
(1+\beta + i \epsilon )^{1-\sigma}} 
\een
\be\label{c3}
\times \, \left\{ 2^{- 2 i \nu} \, (1-\alpha)^{\frac{1+|n|}{2} + i \, \nu} 
\, (1-\beta + i \epsilon )^{\frac{1-|n|}{2} + i \, \nu} - 1 \right\}.
\ee
The $\beta$-integral in \eq{c3} can be done by distorting the contour
around one of the branch cuts. Performing $\alpha$-integral as well
and taking $\sigma \rightarrow 0$ limit of \eq{c3} yields
\be\label{c5}
C (n, \nu) \, = \, 4 \, \pi \, \chi (n, \nu).
\ee
Combining Eqs. (\ref{eigen}) and (\ref{c5}) with \eq{II} gives us
\eq{goal}, as desired.

\renewcommand{\theequation}{B\arabic{equation}}
  \setcounter{equation}{0}  
  \section*{Appendix B: Three-Gluon Exchange Amplitude}
\label{dec}

Here we decompose the function 
\be
\ln^3 \bigg| \frac{\rho_{11'} \, \rho_{22'}}{\rho_{12'} \, 
\rho_{1'2}} \bigg|
\ee
in the series of \eq{decom}. First we note that
\ben
\nabla_1^2 \, \nabla_2^2 \, 
\ln^3 \bigg| \frac{\rho_{11'} \, \rho_{22'}}{\rho_{12'} \, 
\rho_{1'2}} \bigg| \, = \, 6 \, \pi \, \left[ \delta (\rho_{22'}) - 
 \delta (\rho_{21'}) \right] \, \int d^2 \rho_3 \, \frac{|\rho_{1'2'}|^2}{
|\rho_{31'}|^2 \, |\rho_{32'}|^2} \, \left[ 2 \, \delta (\rho_{31}) - 
 \delta (\rho_{11'}) \right] 
\een
\be\label{st1}
+  6 \, \pi \, \left[ \delta (\rho_{11'}) - 
 \delta (\rho_{12'}) \right] \, \int d^2 \rho_3 \, \frac{|\rho_{1'2'}|^2}{
|\rho_{31'}|^2 \, |\rho_{32'}|^2} \, \left[ 2 \, \delta (\rho_{32}) - 
 \delta (\rho_{22'}) \right]. 
\ee
Then, employing (see Eq. (25) in \cite{Lipatov1})
\be
(2 \pi)^4 \, \delta (\rho_{11'}) \, \delta (\rho_{22'}) \, = \, 
\sum_{n=-\infty}^\infty \, \int_{-\infty}^\infty d \nu \, \int d^2
\rho_0 \, \frac{16 \left(\nu^2 + \frac{n^2}{4}\right)}{|\rho_{12}|^2 \, 
|\rho_{1'2'}|^2} \, E^{n, \nu} (\rho_{10}, \rho_{20}) \, 
E^{n, \nu \, *} (\rho_{1'0}, \rho_{2'0})
\ee
in \eq{st1} and remembering that (see Eqs. (21) of \cite{Lipatov1})
\be
\nabla_1^2 \, \nabla_2^2 \, E^{n, \nu} (\rho_{10}, \rho_{20}) \, = \, 
\frac{16}{|\rho_{12}|^4} \, \left[ \nu^2 + \frac{(n+1)^2}{4} 
\right] \, \left[ \nu^2 + \frac{(n-1)^2}{4} \right] \, 
E^{n, \nu} (\rho_{10}, \rho_{20})
\ee
yields
\ben
\ln^3 \bigg| \frac{\rho_{11'} \, \rho_{22'}}{\rho_{12'} \, 
\rho_{1'2}} \bigg| \, = \, \frac{3}{2 \, \pi^3} \, \sum_{\mbox{odd n}} 
\, \int_{-\infty}^\infty d \nu \, \int d^2
\rho_0 \, \frac{\nu^2 + \frac{n^2}{4}}{\left[ \nu^2 + \frac{(n+1)^2}{4} 
\right] \, \left[ \nu^2 + \frac{(n-1)^2}{4} \, \right]} \, 
E^{n, \nu} (\rho_{10}, \rho_{20})
\een
\be\label{st2}
\times \, \int d^2 \rho_3 \, \frac{|\rho_{1'2'}|^2}{
|\rho_{31'}|^2 \, |\rho_{32'}|^2} \, \left[ E^{n, \nu \, *} (\rho_{1'0},
\rho_{30}) +  E^{n, \nu \, *} (\rho_{30}, \rho_{2'0}) -  
E^{n, \nu \, *} (\rho_{1'0}, \rho_{2' 0}) \right]
\ee
Using \eq{goal} to evaluate \eq{st2} gives the final answer
\ben
\ln^3 \bigg| \frac{\rho_{11'} \, \rho_{22'}}{\rho_{12'} \, 
\rho_{1'2}} \bigg| \, = \, \frac{6}{\pi^2} \, \sum_{\mbox{odd n}} 
\, \int_{-\infty}^\infty d \nu \, \int d^2
\rho_0 \, \frac{\nu^2 + \frac{n^2}{4}}{\left[ \nu^2 + \frac{(n+1)^2}{4} 
\right] \, \left[ \nu^2 + \frac{(n-1)^2}{4} \, \right]} \,
\een
\be\label{st3}
\times \, \chi (n, \nu) \,
E^{n, \nu} (\rho_{10}, \rho_{20}) \, E^{n, \nu \, *} (\rho_{1'0},
\rho_{2' 0}).
\ee

\end{document}